\documentclass[letterpaper,comsoc]{IEEEtran}
\usepackage{cite}
\usepackage{algorithm}
\usepackage{amsmath,amssymb,amsfonts}
\usepackage{algorithmic}
\usepackage{graphicx}
\usepackage{textcomp}
\usepackage{epstopdf}
\usepackage{stfloats}
\usepackage{setspace}
\ifCLASSOPTIONcompsoc
\usepackage[caption=false, font=normalsize, labelfont=sf, textfont=sf]{subfig}
\else
\usepackage[caption=false, font=footnotesize]{subfig}
\usepackage{bm}

\usepackage[T1]{fontenc}% optional T1 font encoding
\usepackage{cite}
\usepackage{amsmath,amssymb,amsfonts}
\usepackage{algorithmic}
\usepackage{graphicx}
\usepackage{textcomp}
\usepackage{xcolor}
\usepackage{bm}
\usepackage{changepage}
\usepackage{amsmath}
\usepackage{amstext}
\usepackage{stfloats}
\usepackage{url}

\usepackage{amsmath}
\begin{document}

\title{
%When PASS Meets OFDM: Modeling and Analysis for Multi-Carrier Pinching-Antenna Systems
%OFDM PASS: Frequency-Selective Modeling and Multi-Carrier Performance Analysis for Pinching-Antenna Systems
\huge Frequency-Selective Modeling and Analysis for OFDM-Integrated Wideband Pinching-Antenna Systems 
	\thanks{
%	\emph{Corresponding author: Ji Wang}
	
	Jian Xiao and Ji Wang are with the Department of Electronics and Information Engineering, College of Physical Science and Technology, Central China Normal University, Wuhan 430079, China (e-mail: jianx@mails.ccnu.edu.cn; jiwang@ccnu.edu.cn). 
	
	Ming Zeng is with the Department of Electric and Computer Engineering, Laval University, Quebec City, Canada (email: ming.zeng@gel.ulaval.ca).
	
	Yuanwei Liu is with the Department of Electrical and Electronic Engineering, The University of Hong Kong, Hong Kong (e-mail: yuanwei@hku.hk).
	
	George K. Karagiannidis is with the Department of Electrical and Computer Engineering, Aristotle University of Thessaloniki, 541 24 Thessaloniki, Greece (e-mail: geokarag@auth.gr).
	}
	}
\author{Jian Xiao, Ji Wang, Ming Zeng, Yuanwei Liu,~\IEEEmembership{Fellow,~IEEE}, and George K. Karagiannidis,~\IEEEmembership{Fellow,~IEEE}}% <-this % stops a space
\maketitle
\begin{abstract}
This letter investigates the integration of pinching-antenna systems (PASS) with orthogonal frequency division multiplexing (OFDM) to ensure their compatibility and to explore the frequency-selective behavior inherent to PASS. First, an end-to-end channel model for OFDM PASS is proposed based on electromagnetic-compliant modeling of waveguides and coupled-mode theory, which includes frequency-dependent waveguide {attenuation}, dispersion and antenna coupling effect. Furthermore, a critical dependence of the OFDM cyclic prefix (CP) overhead on the proximity of the operating frequency to the waveguide cutoff is revealed. Moreover, the phase misalignment effect across subcarriers in OFDM PASS is derived for an approximate pinching antenna location strategy based on path loss minimization, which reveals the phase misalignment is exacerbated for wider bandwidths and larger array size. Numerical results show that: 1) frequency-selective effects in OFDM PASS lead to substantial variations in subcarrier achievable rates, highlighting the necessity of operating above the waveguide cutoff frequency for effective communications; 2) waveguide dispersion mandates considerable CP overhead when operating near the cutoff frequency, severely impacting the spectral efficiency of OFDM PASS; and {3) the gentle linear waveguide attenuation in a practical PASS significantly more advantageous than the severe logarithmic path loss characteristic of fixed-location antennas.}

\end{abstract}

\IEEEpeerreviewmaketitle

\section{Introduction}
\IEEEPARstart{T}{he} evolution of wireless communication systems has been driven by advances in antenna technology, from traditional multiple-input multiple-output (MIMO) systems to emerging flexible MIMO paradigms \cite{9770295}. While typical flexible MIMO technologies, such as fluid and movable antennas, offer improved adaptability, they can still face significant challenges in robustly mitigating severe large-scale fading. This is particularly acute in high-frequency communication bands, where signals inherently suffer from significant propagation attenuation and are highly susceptible to line-of-sight (LoS) obstructions. To address these persistent issues more effectively, pinching-antenna systems (PASS) are proposed to create controllable LoS links by leveraging dynamic antenna repositioning over potentially large distances on the dielectric waveguides, thereby directly mitigating free-space path loss and overcoming LoS blockages \cite{Fukuda2022, Ding2024}. {As PASS is designed for high-frequency bands where vast bandwidth is available for high-rate communication, it is natural to investigate the integration of PASS with the typical Orthogonal Frequency Division Multiplexing (OFDM) technology to ensure the compatibility of PASS and to fully exploit its potential in modern wideband systems.} However, this integration is challenged by the inherently frequency-dependent behavior of waveguide propagation and pinching antenna (PA) coupling.

In PASS, signals propagate with minimal loss through dielectric waveguides and are radiated into free space via PAs. The coupling mechanisms between waveguides and PAs are governed by coupled mode theory{\cite{collin1990field,104225,Wang2025}}, which determines the efficiency of signal transfer. While waveguides provide low-loss signal propagation, the inherent frequency selectivity of waveguides poses unique challenges for OFDM PASS. In particular, unlike free-space channels, the waveguide acts as a dispersive medium and exhibits significant dispersion, which inherently limits the performance of high-data-rate PASS links by leading to pulse distortion. This limitation becomes particularly pronounced in OFDM PASS, where the propagation constant and group velocity vary nonlinearly with subcarrier frequency. In addition, the coupling efficiency between waveguides and PAs is highly sensitive to frequency-dependent phase matching and coupling strength\cite{pozar2021microwave}. Previous works have investigated single-carrier PASS scenarios \cite{10896748,zeng2025sum,papanikolaou2025resolving}, but a focused, rigorous analysis detailing the interplay of frequency selectivity in OFDM PASS is crucial for understanding system performance and limitations.

To thoroughly elucidate the frequency-selective behavior of PASS, this letter investigates an electromagnetic-compliant transmission modeling and analysis for OFDM PASS. Specifically, we first develop an end-to-end channel model based on waveguide theory and coupled mode principles that captures the frequency-dependent waveguide propagation and PA coupling. By deriving the group delay in waveguide propagation, we establish a critical dependence of the OFDM cyclic prefix (CP) overhead on the proximity of the operating frequency to the waveguide cutoff. Furthermore, we analyze the phase offset effect for a frequency-independent PA location approximation strategy derived from geometric path loss considerations. Numerical results illustrate that there is significant variability in achievable subcarrier rates directly due to frequency selectivity, while significant CP overhead driven by waveguide dispersion near the cutoff critically limits spectral efficiency. {Moreover, the fundamental performance comparison is analyzed between practical OFDM PASS and conventional fixed-location antenna systems, where the linear transmission loss within the practical waveguide is significantly more advantageous than facing the logarithmic increase of free-space path loss in fixed-location antenna systems.}

\section{Transmission Model of OFDM PASS}
\begin{figure}[t]
	%		\centering
	%		\setlength{\belowcaptionskip}{-1.2cm}
	\centerline{\includegraphics[width=3.0in]{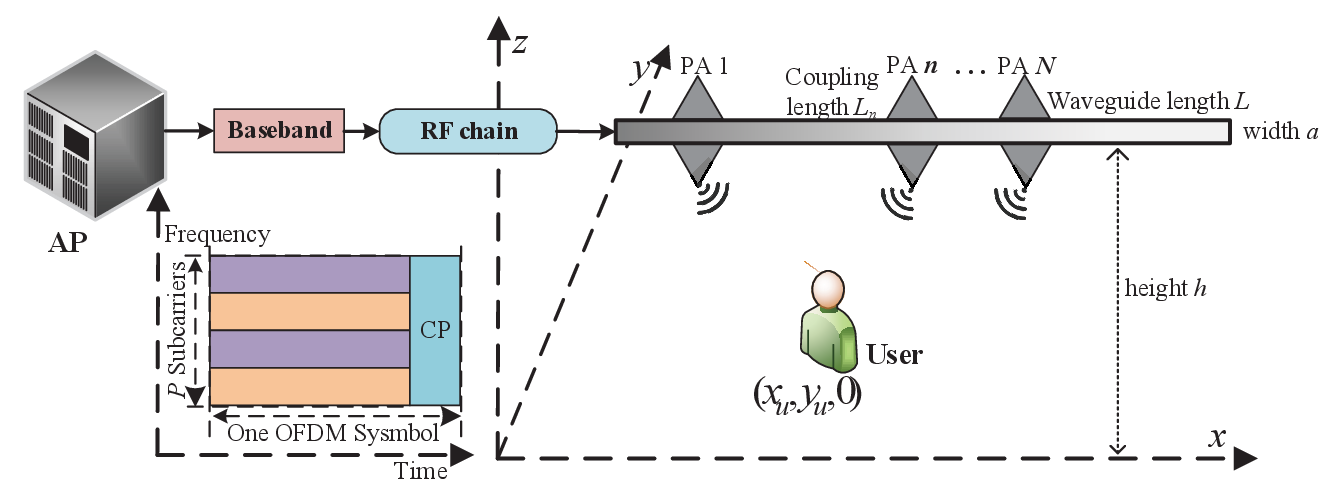}}
	%	\captionsetup{font={small},justification=raggedright,singlelinecheck=off}
	\caption{OFDM-integrated pinching-antenna systems.}
	\label{system}
\end{figure}
As shown in Fig.~\ref{system}, we consider a downlink PASS with a single dielectric waveguide placed at height $h$ along the x-axis. An access point (AP) feeds the waveguide at $x=0$. $N$ PAs are located at positions ${x}_n$ ($n=1,\ldots,N$) along the waveguide. The system serves a single-antenna user at ${\boldsymbol{\psi}}_u = (x_u, y_u, 0)$ using OFDM with $P$ subcarriers at frequencies $f_p$ ($p=1,\ldots,P$). In PASS, signal transmission is characterized by three principal stages: initial propagation along the dielectric waveguide, subsequent coupling to the PAs, and finally, free-space radiation from the PAs to the user.
\subsection{Modeling Frequency-Selective Propagation in Waveguides}

\subsubsection{Cutoff Frequency and Basic Propagation}
Waveguides support distinct propagation modes, e.g., transverse electric (TE) or transverse magnetic (TM) modes \cite{collin1990field, pozar2021microwave}, each characterized by a cutoff frequency $f_0 = \frac{k_c c}{2\pi}$, where $c$ denotes the speed of light, and $k_c$ is the cutoff wavenumber\footnote{While the synthesized waveguide for all-pass frequency band has been investigated in ultrafast electronics \cite{2023Synthesized}, the $\rm{TE}_{10}$ mode is expected to be adopted for the rectangular waveguide in PASS as it is typically the dominant mode, characterized by the lowest cutoff frequency and a stable field configuration for efficient coupling with PAs.}. Signals with frequencies $f_p < f_0$ are evanescent and decay rapidly, while those with $f_p > f_0$ propagate. For the dominant $\mathrm{TE}_{10}$ mode in a rectangular waveguide of width $a$, the cutoff frequency can be simplified as $f_0 = c/(2a)$ \cite{pozar2021microwave}. 

\subsubsection{Transfer Function, Dispersion, and Group Delay}
The transmission of a signal at subcarrier $f_p$ over a distance $x_n$ in the waveguide is described by the transfer function $T(f_p) = e^{-\gamma(f_p) x_n}$ \cite[Ch. 3]{collin1990field}. The complex propagation constant $\gamma(f_p) = \alpha_{\rm{g}}(f_p) + j \beta_{\rm{g}}(f_p)$ includes the waveguide attenuation coefficient $\alpha_{\rm{g}}(f_p)$ in Nepers per meter (Np/m) and the phase constant $\beta_{\rm{g}}(f_p)$. {Specifically, the attenuation constant $\alpha_{\rm{g}}(f_p)$ mainly originating from the conductor loss due to the finite conductivity of the waveguide wall and dielectric loss due to the loss angle of the dielectric. A simplified frequency-dependent $\alpha_{\rm{g}}(f_p)$ is given by $\alpha_{\rm{g}}(f_p) = C_1\sqrt{f_p} + C_2f_p$, where $C_1$ and $C_2$ represent the coefficient for conductor loss and dielectric loss, respectively.} For the $\mathrm{TE}_{10}$ mode ($f_p > f_0$), the phase constant is $\beta_{\rm{g}}(f_p) = \sqrt{(2\pi f_p/c)^2 - (\pi/a)^2}$ \cite{collin1990field}. If $f_p < f_0$, $\beta_{\rm{g}}(f_p)$ becomes imaginary, leading to high {signal} attenuation and $T(f_p) \approx 0$.

The frequency dependence of $\beta_{\rm{g}}(f_p)$ causes waveguide dispersion, resulting in a frequency-dependent group velocity that governs the propagation of signal energy and can be expressed as \cite{pozar2021microwave}
\begin{equation}
\begin{split}\label{vg_condensed} % Suggestion: Unique label
v_{\rm{g}}(f_p) = \left( \frac{d \beta_{\rm{g}}(f_p)}{d {f_p}} \right)^{-1}= c \sqrt{1 - \left( \frac{f_{0}}{f_p} \right)^2}.
\end{split}
\end{equation}
Hence, the propagation delay for a subcarrier $f_p$ over distance $x_n$ is also frequency-dependent, which can be expressed as
\begin{equation}
\begin{split}\label{tau_g_condensed} % Suggestion: Unique label
\tau_{\rm{g}}(f_p) = \frac{x_n}{v_{\rm{g}}(f_p)}.
\end{split}
\end{equation}
This subcarrier-specific delay $\tau_{\rm{g}}(f_p)$ is a critical aspect of OFDM PASS, as varying delays across subcarriers can induce inter-symbol interference (ISI).

\subsection{Waveguide-PA Coupling Mechanisms}
When the guided wave interacts with PAs, the resulting signal coupling principle can be described by the coupled-mode theory \cite{collin1990field, Wang2025, 104225}. The coupling efficiency of PA $n$ at subcarrier frequency $f_p$ depends on the {couping} length $L_n$ of PA $n$ on the waveguide, the frequency-dependent coupling strength $\kappa(f_p)$, and the phase mismatch $\Delta\beta(f_p) = \beta_{\rm{g}}(f_p) - \beta_{\rm{p}}(f_p)$. Here, 
$\beta_{\rm{p}}(f_p)$ is the propagation constant of PAs and is given by
\begin{equation}
\begin{split}\label{eq_beta_p} % Suggestion: Unique label, e.g., eq:beta_p
\beta_{\rm{p}}(f_p) = \frac{2\pi f_p}{c} n_{\rm{p}}(f_p),
\end{split}
\end{equation}
where $n_{\rm{p}}(f_p)$ is the effective refractive index of PAs. Note that $\beta_{\rm{p}}(f_{p})$ is governed by $n_{\rm{p}}(f_p)$, which is related to the material structure of PAs, while $\beta_{\rm{g}}(f_p)$ is dictated by physical dimensions of waveguides and cutoff frequency $f_0$, thereby leading to a non-linear relationship with frequency \cite{pozar2021microwave}.

Consequently, the distinct dispersion characteristics of the waveguide mode $\beta_{\rm{g}}(f_p)$ and the PA mode $\beta_{\rm{p}}(f_p)$ make achieving perfect phase matching ($\Delta\beta(f_p)=0$) simultaneously across all subcarriers generally infeasible in OFDM PASS. The inherently frequency-dependent $\Delta\beta(f_p)$ becomes a primary source of coupling efficiency variation over the band. While idealized $\Delta\beta=0$ in single-carrier scenarios have been analyzed \cite{Wang2025}, this work focuses on the more general and practical frequency-selective case where $\Delta\beta(f_p)$ varies. Let $A(x)$ and $B(x)$ be the complex amplitudes in the waveguide and PA, respectively, along the coupling length $x \in [0, L_n]$. Their evolution is described by the coupled-mode equations:
\begin{equation}
\begin{split}\label{eq_coupled_mode_eqs} % Suggestion: Unique label
\frac{dA(x)}{dx} = -j\kappa(f_p) B(x)e^{-j\Delta\beta(f_p) x},\\
\frac{dB(x)}{dx} = -j\kappa(f_p) A(x)e^{j\Delta\beta(f_p) x}.
\end{split}
\end{equation}
With initial conditions $A(0)=1$ (normalized waveguide input) and $B(0)=0$, the solutions are
\begin{equation}
\begin{split}\label{eq_coupled_mode_solutions} % Suggestion: Unique label
&A(x) = \left( \cos(S_p x) + j\frac{\Delta\beta(f_p)/2}{S_p}\sin(S_p x) \right) e^{-j\Delta\beta(f_p) x / 2},\\
&B(x) = \left( -j \frac{\kappa(f_p)}{S_p}\sin(S_p x) \right) e^{j\Delta\beta(f_p) x / 2},
\end{split}
\end{equation}
where $S_p = \sqrt{\kappa(f_p)^2 + (\Delta\beta(f_p)/2)^2}$ is a interaction parameter.

\textbf{Definition 1:} The local complex coupling factor for PA $n$ with length $L_n$ at subcarrier $p$ is defined as
\begin{equation}
\begin{split}\label{eq_alpha_prime} % Suggestion: Unique label
\alpha'_{{\rm{PA}},n}(f_p)&= {B(L_n)|_{A(0)=1, B(0)=0}}\\
&= -j \frac{\kappa(f_p)}{S_p}\sin(S_p L_n) e^{j\Delta\beta(f_p) L_n / 2}.
\end{split}
\end{equation}

The complex amplitude ratio remaining in the waveguide after PA $n$ with length $L_n$ can be expressed as
\begin{equation}
\begin{split}\label{eq_R_n} % Suggestion: Unique label
&{\alpha'_{{\rm{WG}},n}}(f_p) = A(L_n)|_{A(0)=1, B(0)=0}\\
&=\left( \cos(S_{p,n} L_n) + j\frac{\Delta\beta_n(f_p)/2}{S_{p,n}}\sin(S_{p,n} L_n) \right) e^{-j\Delta\beta_n(f_p) L_n / 2}.
\end{split}
\end{equation}

Let $E_{\text{in},1}(f_p)$ be the complex amplitude at the waveguide input. Neglecting inter-PA propagation effects for this cascaded coupling analysis, the amplitude entering PA $n$ is $E_{\text{in},n}(f_p) = E_{\text{in},1}(f_p) \prod_{i=1}^{n-1} \alpha'_{{\rm{WG}},i}(f_p)$. The amplitude coupled out by PA $n$ is thus $E_{\text{coup},n}(f_p) = E_{\text{in},n}(f_p) \alpha'_{{\rm{PA}},n}(f_p)$.

\textbf{Definition 2:} The overall effective complex coupling factor for PA $n$ at subcarrier $p$ is defined as 
\begin{equation}
\begin{split}\label{eq_alpha_double_prime} % Suggestion: Unique label
\alpha''_{n,p}(f_p) &= E_{\text{coup},n}(f_p) / E_{\text{in},1}(f_p)\\
&=\left( \prod_{i=1}^{n-1} \alpha'_{{\rm{WG}},i}(f_p) \right) \times \alpha'_{{\rm{PA}},n}(f_p).
\end{split}
\end{equation}
The factor $\alpha''_{n,p}(f_p)$ encapsulates the frequency-selective coupling to PA $n$, considering the influence of all preceding PAs.

\subsection{End-to-End Effective Channel Gain of OFDM PASS}

Let $d_{u,n} = ||\boldsymbol{\psi}_u - ({x}_n, 0, h)||$ denote the distance between the user and PA $n$. Assuming isotropic PA, the free-space channel channel ${h}_{n,p}$ between the user and PA $n$ can be expressed as 
\begin{equation}
\begin{split}\label{G}
{h}_{n,p} = \frac{c}{4\pi f_p d_{u,n}}e^{-j\frac{2\pi f_p}{c} d_{u,n}}.
\end{split}
\end{equation}

The end-to-end effective channel gain for subcarrier $p$ via PA $n$ can be expressed as
\begin{equation}
\begin{split}\label{E2E}
h_{\text{eff}, n, p}({x}_n) &= T(f_p) \times \alpha''_{n,p}(f_p) \times  {h}_{n,p},
\end{split}
\end{equation}
where $T(f_p)$, $\alpha''_{n,p}(f_p)$ and ${h}_{n,p}$ represent the in-waveguide propagation, the waveguide-PA coupling,  the free-space propagation, respectively\footnote{{For the case of uplink PASS, the bidirectional power split at coupling and cumulative leakage during in-waveguide propagation should be considered. Hence, the overall end-to-end uplink channel differs from the downlink channel. However, the fundamental frequency dependent behavior of PASS analyzed in this letter, such as waveguide dispersion and frequency-selective fading, are equally critical in the uplink.}}.
The total effective channel gain for subcarrier $p$ is the coherent sum from PA locations $\mathbf{x}=[{x}_1, \ldots, {x}_N]$ and can be expressed as
\begin{equation}
\begin{split}\label{G}
H_p(\mathbf{x}) = \sum_{n=1}^N h_{\text{eff}, n, p}({x}_n) .
\end{split}
\end{equation}

Suppose AP transmits symbol $s_p$ with power $P_p$ on subcarrier $p$, the received signal at the user over subcarrier $p$ can be expressed as
\begin{equation}
\begin{split}\label{G}
y_p = H_p(\mathbf{x}) \sqrt{P_p} s_p + z_p,
\end{split}
\end{equation}
where \(z_p \sim \mathcal{CN}(0, \sigma^2)\) denotes Gaussian noise. The received signal-to-noise ratio (SNR) can be expressed as
\begin{equation}
\begin{split}\label{G}
\text{SNR}_p(\mathbf{x}) = \frac{P_p |H_p(\mathbf{x})|^2}{\sigma^2}.
\end{split}
\end{equation}
The achievable rate across all \( P \) subcarriers is thus given by 
\begin{equation}
\begin{split}
R_{\text{total}} = {\frac{B}{P + L_{\rm{CP}}}}\sum_{p=1}^{P} R_p,
\end{split}
\end{equation}
where $R_p = \log_2 \left(1 + \text{SNR}_p(\mathbf{x})  \right)$ denotes achievable rate over subcarrier \( p \). Parameters $B$ and $L_{\rm{CP}}$ denote the bandwidth and the length of CP in OFDM PASS, respectively.

\section{Performance Analysis in OFDM PASS}
\subsection{CP Requirement in OFDM PASS}
To prevent ISI in OFDM PASS, the OFDM CP length in time, $T_{\rm{CP}}$, must exceed the maximum channel delay spread $\Delta\tau_{\rm{total}}$, which arises from two primary sources: (i) waveguide dispersion over the array length $L_{\rm{array}}= {x}_N - {x}_1$, and (ii) the spatial-wideband effect in free-space propagation. Specifically, in waveguide propagation, the maximum differential delay across the array aperture $L_{\rm{array}}$ and bandwidth $B$ is given by
\begin{equation}
\begin{split}\label{G}
\Delta \tau_{{\rm{g}},\rm{max}} &\approx \max_p \left|\frac{d\beta_{\rm{g}}(f_p)}{d\omega_p}\right| L_{\rm{array}} - \min_p \left|\frac{d\beta_{\rm{g}}(f_p)}{d\omega_p}\right| L_{\rm{array}} \\
&= \left|\frac{1}{v_{\rm{g}}(f_{\rm{1}})} - \frac{1}{v_{\rm{g}}(f_{P})}\right| L_{\rm{array}},
\end{split}
\end{equation}
where $\omega_p=2\pi f_p$,
$f_{1}$ and $f_{P}$ denote the minimum and maximum subcarrier frequencys in OFDM systems, respectively.

{The free-space propagation delay difference from PAs located at different positions along the waveguide to the user can be expressed as
\begin{equation}
\begin{split}\label{G}
\Delta\tau_{\rm{f,max}} = \frac{\max_n d_{u,n} - \min_n d_{u,n}}{c}.
\end{split}
\end{equation} 
Consequently, a sufficient CP length in time should satisfy $T_{\rm{CP}} > \Delta\tau_{\rm{total}}=\Delta\tau_{\rm{g,max}} + \Delta\tau_{\rm{f,max}}$\footnote{{Note that the waveguide-induced delay spread $\Delta\tau_{\rm{g,max}}$ is frequency-dependent because group velocity $v_{\rm{g}}$ within the dispersive waveguide medium inherently varies with signal frequency $f_p$. In contrast, the free-space array-induced delay $\Delta\tau_{\rm{f,max}}$ is frequency-independent, as the constant speed of light $c$ in the non-dispersive free-space medium. Nevertheless, the geometric delay result in frequency-dependent phase shifts across the wideband signal's subcarriers within $h_{n,p}$, contributing to the overall channel frequency selectivity.}}. In a digital implementation, this CP duration is realized by $L_{\rm{CP}}$ discrete samples, such that $T_{\rm{CP}} = L_{\rm{CP}} \times T_s$, where $T_s$ is the sampling period. Given the sampling period $T_s = 1/B$, the number of CP samples must therefore satisfy $L_{\rm{CP}} > \Delta\tau_{\rm{total}} \times B$.} Hence, the significant waveguide dispersion and large array apertures will result in a greater $L_{\rm{CP}}$, reducing the overall system spectral efficiency.

\subsection{Phase Misalignment Effect for Frequency-Independent PA Location Approximation}
{According to the frequency-selective channel model established in Eq.~\eqref{E2E}, deriving a general closed-form solution for the optimal PA locations is intractable to maximize the achievable rate of OFDM PASS, which requires solving a non-convex optimization problem. In particular, the waveguide attenuation $\alpha_\text{g}$ and the cascaded coupling factor $\alpha''_{n,p}$ favor placing PAs closer to the AP to maximize available signal power \cite{Xu2025sum}, whereas minimizing the geometric path loss ${h}_{n,p}$ in free-space favors placing PAs closer to the user. For the purpose of tractable performance analysis and to establish a baseline for evaluating the impact of these frequency-selective phenomena, we adopt a simplified and idealized antenna placement model that focuses solely on frequency-independent geometric considerations \cite{10896748}.} Considering a ideal single-carrier PASS at the center frequency $f_c$, PA locations $\tilde{\mathbf{x}}=[\tilde{{x}}_1,\ldots,\tilde{{x}}_N]$ can be optimized to maximize the frequency-independent geometric path gain $f(\tilde{\mathbf{x}}) = \sum_{n=1}^{N} [(\tilde{x}_n - x_u)^2 + C]^{-\frac{1}{2}}$, where $C=y_u^2+h^2$, subject to minimum spacing $\tilde{x}_n - \tilde{x}_{n-1} \ge \Delta= L_n+G$. Here, $G$ denotes the minimum physical spacing of adjacent PAs\footnote{In scenarios where achieving sufficient power coupling necessitates, a relatively large $L_n$, e.g., due to a weak coupling coefficient $\kappa$, it is possible that this minimum physical spacing $L_n+G > \lambda/2$. The conventional $\lambda/2$ spacing guideline, often employed in array design to suppress grating lobes, cannot be satisfied. However, PASS systems typically establish desired links by optimizing PA locations $\tilde{\mathbf{x}}$ along the waveguide, often focusing on strengthening a LoS path towards a specific user rather than achieving wide-angle electronic scanning via phase shifters.}. Under the assumption that the objective function $f(\tilde{\mathbf{x}})$ restricted by $\tilde{x}_n = \tilde{x}_1 + (n-1)\Delta$ is unimodal with respect to $\tilde{x}_1$, which holds if $C \ge (N-1)^2\Delta^2$, the optimal solution in downlink PASS is given by\cite{10896748}
\begin{equation}
\begin{split}\label{eq:solution_approx} % 请确保标签唯一
\tilde{x}_n = x_u + \left(n - 1 - \frac{N-1}{2}\right)\Delta, \quad n=1, ..., N.
\end{split}
\end{equation}

%This solution, $\tilde{\mathbf{x}}$, solely optimizes frequency-independent geometric path loss and critically ignores signal phase.
 In OFDM PASS, since the total phase experienced by the signal from each PA is inherently frequency-dependent due to waveguide dispersion, coupling, and free-space propagation, employing the frequency-independent locations $\tilde{\mathbf{x}}$ results in varying inter-PA phase differences for different subcarriers $p$. Consequently, the condition for constructive interference, i.e., phase differences being integer multiples of $2\pi$, cannot be simultaneously met across all frequencies. This inevitable phase misalignment causes signals on some subcarriers to undergo non-coherent or even destructive interference.

\subsection{Phase Misalignment Analysis of OFDM PASS}
To characterize the detrimental effect of this phase misalignment introduced by the single-carrier approximation, we analyze the magnitude of the phase variation itself. We aim to find the phase of the signal component arriving at the user via PA $n$, relative to the phase of $E_{\rm{in},1}(f_p)$ at the input of the waveguide. The complex amplitude $E_{n,p}$ of the signal component at the user can be expressed as
\begin{equation}
\begin{split}\label{G}
 E_{n,p} = & E_{\rm{in},1}(f_p) \times T(f_p) \times \left( \prod_{i=1}^{n-1} \alpha'_{{\rm{WG}},i}(f_p) \right) \\
 & \times \alpha'_{{\rm{PA}},n}(f_p) \times h_{n,p} \times h_{\rm{ant}},
 \end{split}
\end{equation}
where $h_{\rm{ant}}$ represents other constant phase shifts or gains from transmit and receive antennas. The total phase of $E_{n,p}$ relative to $E_{\rm{in},1}(f_p)$ is given by
\begin{equation}
\begin{split}\label{G}
&\arg(E_{n,p}/E_{\rm{in},1}(f_p)) \\
 &= -\beta_{\rm{g}}(f_p)\tilde{x}_n + \phi_{{\rm{accum}},n-1,p} + \phi_{{\rm{coup}},n,p} - \phi_{n,p} + \phi_{\rm{const}},
\end{split}
\end{equation}
where $\phi_{{\rm{accum}},n-1,p} = \sum_{i=1}^{n-1} \arg(\alpha'_{{\rm{WG}},i}(f_p))$, $\phi_{{\rm{coup}},n,p} = \arg(\alpha'_{{\rm{PA}},n}(f_p)) = \frac{\Delta\beta_n(f_p) L_n}{2} - \frac{\pi}{2}$ $\pm \pi$, $\phi_{n,p} = \arg(h_{n,p}) =\frac{2\pi f_p}{c}d_{u,n}$ and $\phi_{\rm{const}}=\arg(h_{\rm{ant}})$ denote the phase contribution of corresponding channel components, respectively.

{Defining} the total accumulated phase delay $\Phi_{n,p}(\tilde{\mathbf{x}})$ such that $E_{n,p} \propto e^{-j \Phi_{n,p}(\tilde{x}_n)}$, we have
\begin{equation}
\begin{split}\label{phase}
\Phi_{n,p}(\tilde{\mathbf{x}})
= &\beta_{\rm{g}}(f_p)\tilde{x}_n - \sum_{i=1}^{n-1} \arg\left(\alpha'_{{\rm{WG}},i}(f_p)\right) \\
&- \left(\frac{\Delta\beta_n(f_p) L_n}{2} - \frac{\pi}{2}\right)
+ \frac{2\pi f_p}{c} d_{u,n}(\tilde{x}_n) - \phi_{\rm{const}}.
\end{split}
\end{equation}

\textbf{Lemma 1:} Given $|f_p - f_c| \ll f_c$ in mmWave OFDM PASS, the phase difference between adjacent PAs $\Delta\Phi_{n,p}(\tilde{\mathbf{x}}) = \Phi_{n,p}(\tilde{\mathbf{x}}) - \Phi_{n-1,p}(\tilde{\mathbf{x}})$ can be approximated by
\begin{equation}
\begin{split}\label{phase_error}
\Delta\Phi_{n,p}(\tilde{\mathbf{x}}) \approx \left[ \frac{2\pi \Delta}{v_{\rm{g}}(f_c)} + \frac{2\pi}{c} \Delta_{d,n} \right] (f_p - f_c) +\Delta\Phi_{n,c}(\tilde{\mathbf{x}}),
\end{split}
\end{equation}
where $\Delta_{d,n}={d}_{u,n}-{d}_{u, n-1}$ and $\Delta\Phi_{n,c}$ denotes inter-PA phase differences at the center frequency $f_c$.

\emph{Proof:}
Considering that the derivative of $\arg({\alpha'_{{\rm{WG}},n}}(f_p) )$ contributes less significantly to the linear term than the derivatives of $\beta_{\rm{g}}(f_p)\Delta$ and $\phi_{n,p} - \phi_{n-1,p}$, the accumulated phase term $-\sum_{i=1}^{n-1} \arg(\alpha'_{{\rm{WG}},i}(f_p))$ is omitted. Furthermore, neglecting constant phase terms $\phi_{\rm{const}}$ and the sign difference which cancels in $\Delta\Phi$, the total phase difference experienced by the signal is
$\Delta\Phi_{n,p}(\tilde{\mathbf{x}}) = [\beta_{\rm{g}}(f_p)\tilde{x}_n + \frac{2\pi f_p}{c} {d}_{u,n}] - [\beta_{\rm{g}}(f_p)\tilde{x}_{n-1} + \frac{2\pi f_p}{c} {d}_{u,n-1}] + (\phi_{{\rm{coup}},n,p} - \phi_{{\rm{coup}},n-1,p})= \beta_{\rm{g}}(f_p)(\tilde{x}_n-\tilde{x}_{n-1}) + \frac{2\pi f_p}{c} \Delta_{d,n} + \Delta\phi_{{\rm{coup}}, n, p}$.
Using the first-order Taylor expansion around $f_c (|f_p - f_c| \ll f_c)$:
$\beta_{\rm{g}}(f_p) \approx \beta_{\rm{g}}(f_c) + \beta'_{\rm{g}}(f_c)(f_p - f_c)$
and $\frac{2\pi f_p}{c} {d}_{u,n} \approx \frac{2\pi f_c}{c} {d}_{u,n} + \frac{2\pi {d}_{u,n}}{c}(f_p - f_c)$.
Assuming the differential coupling phase $\Delta\phi_{{\rm{coup}}, n, p} \approx \Delta\phi_{{\rm{coup}}, n, c}$ varies slowly near $f_c$.
Substituting these into the expression for $\Delta\Phi_{n,p}(\tilde{\mathbf{x}})$ and using $\tilde{x}_n-\tilde{x}_{n-1} = \Delta$, we have
\begin{equation}
\begin{split}
 \Delta\Phi_{n,p}(\tilde{\mathbf{x}}) \approx & [\beta_{\rm{g}}(f_c) + \beta'_{\rm{g}}(f_c)(f_p - f_c)]\Delta \\
 &+ \left[\frac{2\pi f_c}{c} \Delta_{d,n}+ \frac{2\pi}{c}\Delta_{d,n}(f_p - f_c)\right] + \Delta\phi_{\text{coup}, n, c}.
\end{split}
\end{equation}
Grouping terms evaluated at $f_c$ gives $\Delta\Phi_{n,c}(\tilde{\mathbf{x}})$, we have
\begin{equation}
\begin{split}
\Delta\Phi_{n,c}(\tilde{\mathbf{x}}) \approx \beta_{\rm{g}}(f_c)\Delta + \frac{2\pi f_c}{c}\Delta_{d,n}+ \Delta\phi_{{\rm{coup}}, n, c}.
\end{split}
\end{equation}
Subtracting this from the expression for $\Delta\Phi_{n,p}(\tilde{\mathbf{x}})$ leaves the terms proportional to $(f_p - f_c)$:
\begin{equation}
\begin{split}
\Delta\Phi_{n,p}(\tilde{\mathbf{x}}) - \Delta\Phi_{n,c}(\tilde{\mathbf{x}}) \approx \left[\beta'_{\rm{g}}(f_c)\Delta + \frac{2\pi}{c}\Delta_{d,n}\right] (f_p - f_c).
\end{split}
\end{equation}
Substituting $\beta'_{\rm{g}}(f_c) = \frac{d\beta_{\rm{g}}}{df}|_{f_c} = \frac{2\pi}{v_{\rm{g}}(f_c)}$, we arrive at \eqref{phase_error}. 
This completes the proof. $\blacksquare$

{\textbf{Lemma 1}} reveals the linear dependence of the phase difference deviation on the frequency offset $(f_p - f_c)$, with the slope determined by the group delay over the PA spacing $\Delta$ and the free-space path difference delay. The approximation $\tilde{x}_n$ works well only when the total phase variation $\Delta\Phi_{n,p} \times B \ll 2\pi$ and the relative bandwidth $B/f_c$ is small. These conditions correspond to narrowband systems, low dispersion ($f_c \gg f_0$), and small array apertures $(N-1)\Delta$.

\begin{figure}[t]
	%		\centering
	%		\setlength{\belowcaptionskip}{-1.2cm}
	\centerline{\includegraphics[width=2.5in]{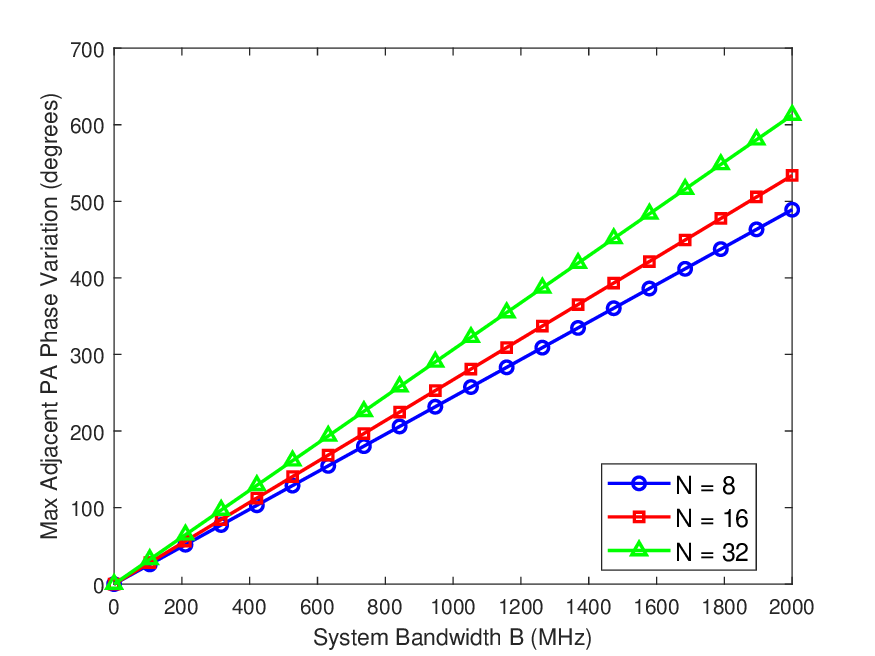}}
	%	\captionsetup{font={small},justification=raggedright,singlelinecheck=off}
	\caption{Phase misalignment analysis of approximate PA locations.}
	\label{error}
\end{figure}

Fig.~\ref{error} investigates the phase consistency for the approximate PA location strategy by analyzing the maximum variation of the phase difference between adjacent PAs, i.e., $\max_{\text{pairs}} |\max_{f_p \in B} (\Delta\Phi_{n,p}) - \min_{f_p \in B} (\Delta\Phi_{n,p})|$, across the system bandwidth $B$ with a mmWave center frequency $f_c=28$ GHz. This metric quantifies the phase swing experienced by the worst-case adjacent PA pair due to frequency variation. The results show that the maximum adjacent PA phase variation increases approximately linearly with the system bandwidth $B$. Furthermore, the phase variation significantly increases with the number of PAs $N$. A larger $N$ implies a larger array aperture, making PASS more sensitive to frequency-dependent phase shifts accumulated over longer waveguide paths and potentially larger variations in $\Delta_{d,n}$ for different PA pairs.

\section{Numerical Results}
In the simulation, unless otherwise stated, we set $N=8$, $P = 64$, $f_c = 28$ GHz, $B = 2$ GHz, $P_p = 30 \text{ dBm}$, $h=5 \text{ m}$, $a = 5.5 \text{ mm}$ and $\boldsymbol{\psi}_u = (5, 2, 0)\text{ m}$. The noise power spectral density is $N_0 = -174 \text{ dBm/Hz}$. {A linear approximation model in the mmWave system is utilized to characterize coupling coefficient $\kappa(f_p) = \kappa_c + \kappa_1 \left( \frac{f_p - f_c}{f_c} \right)$, where the coupling coefficient at the center frequency \( f_c \) is set to \( \kappa_c = 10\) and the slope is set to \( \kappa_1 =5 \). }
For simplicity of hardware implementation, we assume a fixed PA coupling length $L_{\text{PA}} = {\arcsin\left(\sqrt{{1}/{N}}\right)}/{\kappa_c}$ and a frequency-independent refractive index $n_{\rm{p}} = 1.5$.
\begin{figure}[t]
	%		\centering
	%		\setlength{\belowcaptionskip}{-1.2cm}
	\centerline{\includegraphics[width=2.5in]{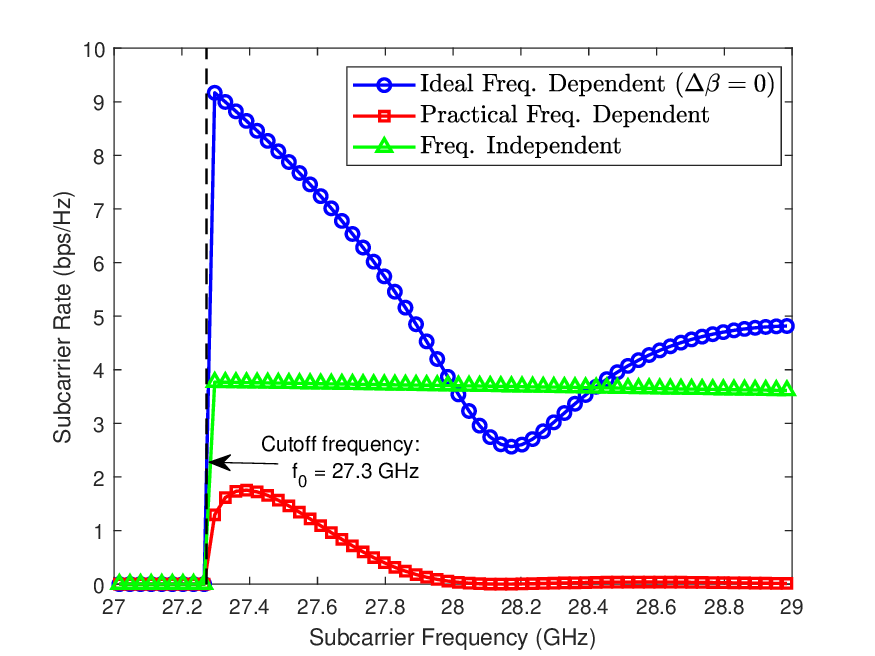}}
	%	\captionsetup{font={small},justification=raggedright,singlelinecheck=off}
	\caption{Achievable rate for different subcarriers of OFDM PASS.}
	\label{rate}
\end{figure}

Fig.~\ref{rate} presents the achievable subcarrier rates for three OFDM PASS setups: (i) an idealized frequency-dependent model with perfect phase match $\Delta\beta=0$, (ii) a practical frequency-dependent model with varying $\Delta\beta(f_p)$, and (iii) a frequency-independent model using parameters from the central frequency $f_c$. A fundamental characteristic is that below cutoff frequency $f_0=c/(2a)=27.3$ GHz, the achievable rate of PASS is negligible due to evanescent waves. {Above $f_0$, the case of perfect phase match $\Delta\beta=0$ represents the ideal performance of OFDM PASS, where rate variations are caused by constructive/destructive interference among the signals from different PAs. However, the practical frequency-dependent model is significantly degraded caused by varying phase mismatch $\Delta\beta(f_p)$, which severely reduces signal coupling efficiency.} Moreover, the frequency-independent model fails to capture the frequency selectivity and provides a misleadingly uniform estimate of OFDM PASS.

\begin{figure}[t]
	%		\centering
	%		\setlength{\belowcaptionskip}{-1.2cm}
	\centerline{\includegraphics[width=2.5in]{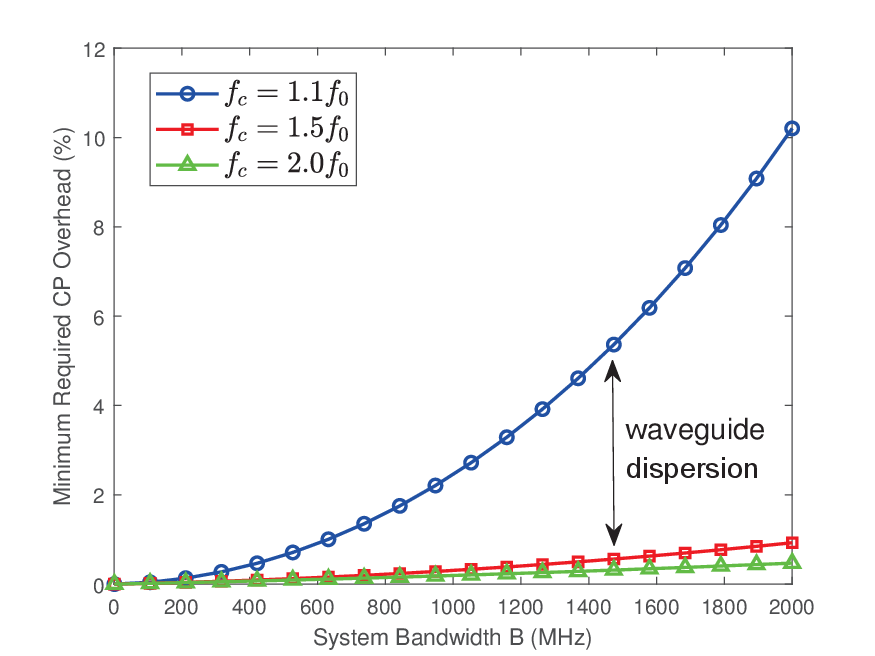}}
	%	\captionsetup{font={small},justification=raggedright,singlelinecheck=off}
	 \caption{Minimum CP overhead of OFDM PASS.}
	\label{cp}
\end{figure}

Fig.~\ref{cp} presents the minimum required CP overhead versus system bandwidth $B$, for different center frequencies $f_c$ relative to the cutoff frequency $f_0$. The CP overhead is calculated to overcome the OFDM PASS delay spread, which is given by $(\Delta\tau_{\rm{total}} / T_{\rm{sym}}) \times 100\%$, where $T_{\rm{sym}} = P/B$ is the useful OFDM symbol duration. This result shows that the required CP overhead increases significantly with system bandwidth $B$. More critically, it shows a strong dependence on the proximity of the operating band to $f_0$. When $f_c$ is close to $f_0$, the waveguide dispersion is severe, resulting in a very large waveguide-induced delay and thus a rapidly increasing CP overhead. Conversely, when $f_c$ is well above $f_0$, the CP overhead is lower, primarily due to the spatial broadband effect in free-space propagation. To maintain spectral efficiency, the operating band of the OFDM PASS should be chosen sufficiently far from the waveguide cutoff frequency.

\begin{figure}[t]
	%		\centering
	%		\setlength{\belowcaptionskip}{-1.2cm}
	\centerline{\includegraphics[width=2.5in]{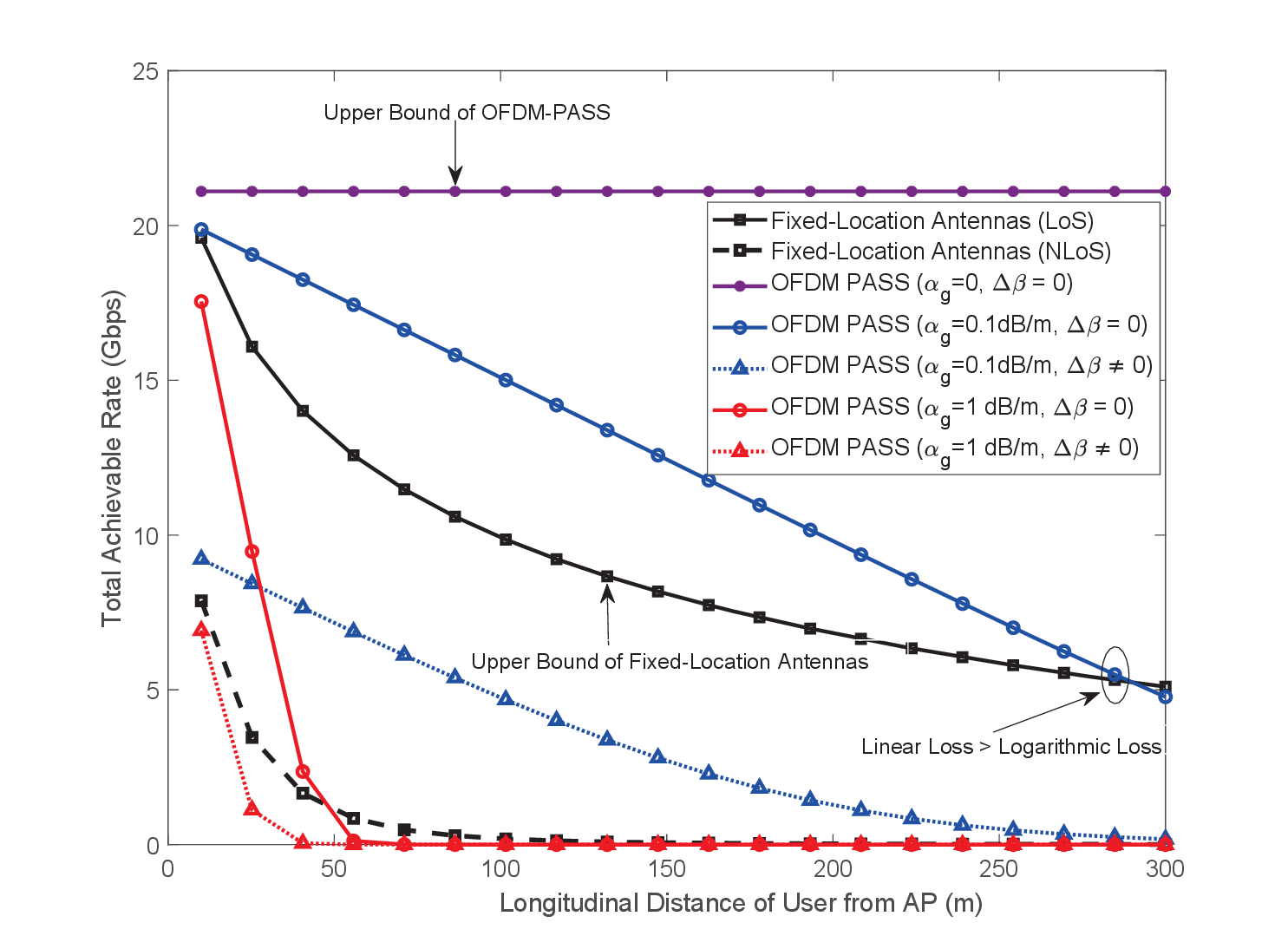}}
	%	\captionsetup{font={small},justification=raggedright,singlelinecheck=off}
	\caption{Performance comparison between PASS and fixed-location antennas.}
	\label{error}
\end{figure}

{Fig.~\ref{error} compares the performance of OFDM PASS with that of conventional fixed-location antennas as a function of the user longitudinal distance from the AP, i.e., the user x-coordinate $x_u$, considering practical waveguide attenuation $\alpha_{\rm{g}}$ at $f_c$ and phase mismatch $\Delta\beta(f_p)$\footnote{{In Section II-A of this letter, the waveguide attenuation $\alpha_{\rm{g}}$ in Np/m describes the exponential decay of a field amplitude over propagation distance.
Considering a more intuitive unit is decibels per meter (dB/m) for characterizing the power ratio of waveguide attenuation, in Fig.~\ref{error}, we utilize a conversion factor to describe $\alpha_\text{g}$ in dB/m, i.e., 1 Np/m $\approx$ 8.686 dB/m. Note that the extended waveguide length is considered in Fig.~\ref{error} primarily to provide insights into the fundamental performance trade-offs of PASS, although such long-distance transmission may not be typical in practical PASS applications.}}. The fixed-location antennas consists of an $N$-antenna array driven by a single radio frequency chain. The rate of the fixed-location LoS system, limited by logarithmic free-space path loss $L_{\text{FS}}(d_{\text{user}}) \propto 20\log_{10}(d_{\text{user}}), (d_{\text{user}}=\sqrt{x_u^2+y_u^2+h^2}$), exhibits a steady decline with the increase of signal propagation distance $x_u$, while its performance of NLoS counterpart collapses due to LoS blockage. In contrast, an ideal OFDM PASS ($\alpha_\text{g}=0$, $\Delta\beta = 0$) maintains a constant and distance-independent performance upper bound. The rate of a more practical PASS is instead dominated by a linear loss $L_{\text{PASS}}(x_u) \propto \alpha_\text{g} \cdot x_u$ (in dB), originating from waveguide attenuation. This fundamental difference in path loss models results in a clear crossover point, up to which the gentle linear loss of low-loss OFDM PASS becomes more advantageous than the severe logarithmic loss of the fixed-location antennas. This demonstrates the superiority of PASS for extending high-throughput coverage to longer distances, although this benefit is highly contingent on using low-loss waveguides and ensuring efficient phase-matched coupling\footnote{{While specialized ultra-low-loss waveguide structures have been developed, e.g., ceramic alumina waveguides whose attenuation factors are less than 10 dB/km in the millimeter-submillimeter band \cite[Ch. 11]{yeh2008essence}, these ultra-low-loss structures often present a conflict with the pinching implementation mechanism of PASS. Their operational principles may rely on rigid mechanical properties that preclude the necessary flexibility for pinching, or on strong electromagnetic field confinement that eliminates the external evanescent field required for interaction. In the first pinching-antenna prototype developed by NTT DOCOMO in 2022 \cite{Fukuda2022}, the common solid dielectric waveguides are used for supporting bended or pinching operations, i.e., PolyTetraFluoroEthylene (PTFE). Therefore, designing novel waveguides that achieve an optimal trade-off between low-loss propagation and the pinching requirements is important.}}. This demonstrates the distinct advantage of PASS in extending high-throughput coverage, particularly in NLoS environments where it maintains a robust link.}

\section{Conclusion}
In this letter, an electromagnetic-compliant end-to-end transmission model of OFDM PASS was established that accounts for frequency-dependent waveguide dispersion and PA coupling. Analysis of this model revealed that the interplay of these frequency-dependent effects is fundamental to system performance, inducing substantial subcarrier-specific gain variations and making CP overhead strongly dependent on proximity to the waveguide cutoff. We further demonstrated that simplified PA placement strategies introduce significant broadband phase shift, limiting their utility to narrowband or low dispersion scenarios. These results underscore the need for frequency-aware PASS optimization, with future work exploring robust PA optimization for multicarrier effects and detailed directional PA coupling characteristics.

\ifCLASSOPTIONcaptionsoff
  \newpage
\fi

\bibliographystyle{IEEEtran}
\bibliography{IEEEabrv,refs_ofdm.bib}
\end{document}